\documentclass[12pt]{article} 
\usepackage{epsfig,graphicx}
\usepackage{fpcp04}
\usepackage{amsmath}
\usepackage{amssymb}
\usepackage{array}
\usepackage{ulem}
\usepackage{wrapfig}

\def\beq{\begin{equation}}
\def\eeq#1{\label{#1}\end{equation}}
\def\eeqn{\end{equation}}

\def\beqa{\begin{eqnarray}}
\def\eeqa#1{\label{#1}\end{eqnarray}}
\def\eeqan{\end{eqnarray}}

\def\overbar#1{\overline{#1}}

\let\bar=\overbar

\def\vev#1{\langle #1 \rangle}

\def\Dslash{\not{\hbox{\kern-4pt $D$}}}
\def\dslash{\not{\hbox{\kern-2pt $\del$}}}

\def\msb{{\bar{\ssstyle M \kern -1pt S}}}

\def\vckm{V_{\rm CKM}}

\def\BB0bar{B^0 {\overline B}^0}
\def\BB0dbar{B_d^0 {\overline B}_d^0}
\def\BB0sbar{B_s^0 {\overline B}_s^0}

\RequirePackage{xspace}

\usepackage{relsize}
\def\babar{\mbox{\slshape B\kern-0.1em{\smaller A}\kern-0.1em
    B\kern-0.1em{\smaller A\kern-0.2em R}}}

\def\d     {\ensuremath{d}\xspace}

\def\Kbar  {\kern 0.2em\overline{\kern -0.2em K}{}\xspace}

\def\Kz    {\ensuremath{K^0}\xspace}
\def\Kzb   {\ensuremath{\Kbar^0}\xspace}
\def\KzKzb {\ensuremath{\Kz \kern -0.16em \Kzb}\xspace}
\def\Kp    {\ensuremath{K^+}\xspace}
\def\Km    {\ensuremath{K^-}\xspace}

\def\KpKm  {\ensuremath{\Kp \kern -0.16em \Km}\xspace}

\def\Dbar    {\kern 0.2em\overline{\kern -0.2em D}{}\xspace}

\def\Dz      {\ensuremath{D^0}\xspace}
\def\Dzb     {\ensuremath{\Dbar^0}\xspace}
\def\DzDzb   {\ensuremath{\Dz {\kern -0.16em \Dzb}}\xspace}
\def\Dp      {\ensuremath{D^+}\xspace}
\def\Dm      {\ensuremath{D^-}\xspace}

\def\DpDm    {\ensuremath{\Dp {\kern -0.16em \Dm}}\xspace}

\def\Bbar    {\kern 0.18em\overline{\kern -0.18em B}{}\xspace}

\def\BB      {\ensuremath{B\Bbar}\xspace} 
\def\Bz      {\ensuremath{B^0}\xspace}
\def\Bzb     {\ensuremath{\Bbar^0}\xspace}
\def\BzBzb   {\ensuremath{\Bz {\kern -0.16em \Bzb}}\xspace}
\def\Bu      {\ensuremath{B^+}\xspace}
\def\Bub     {\ensuremath{B^-}\xspace}

\def\BpBm    {\ensuremath{\Bu {\kern -0.16em \Bub}}\xspace}

\mathchardef\Upsilon="7107
\def\Y#1S{\ensuremath{\Upsilon{(#1S)}}\xspace}

\mathchardef\Deltares="7101
\mathchardef\Xi="7104
\mathchardef\Lambda="7103
\mathchardef\Sigma="7106
\mathchardef\Omega="710A

\def\Deltabar{\kern 0.25em\overline{\kern -0.25em \Deltares}{}\xspace}
\def\Lbar{\kern 0.2em\overline{\kern -0.2em\Lambda\kern 0.05em}\kern-0.05em{}\xspace}
\def\Sigbar{\kern 0.2em\overline{\kern -0.2em \Sigma}{}\xspace}
\def\Xibar{\kern 0.2em\overline{\kern -0.2em \Xi}{}\xspace}
\def\Obar{\kern 0.2em\overline{\kern -0.2em \Omega}{}\xspace}
\def\Nbar{\kern 0.2em\overline{\kern -0.2em N}{}\xspace}
\def\Xb{\kern 0.2em\overline{\kern -0.2em X}{}\xspace}

\newcommand{\tev}{\ensuremath{\mathrm{\,Te\kern -0.1em V}}\xspace}
\newcommand{\gev}{\ensuremath{\mathrm{\,Ge\kern -0.1em V}}\xspace}
\newcommand{\mev}{\ensuremath{\mathrm{\,Me\kern -0.1em V}}\xspace}
\newcommand{\kev}{\ensuremath{\mathrm{\,ke\kern -0.1em V}}\xspace}
\newcommand{\ev}{\ensuremath{\mathrm{\,e\kern -0.1em V}}\xspace}
\newcommand{\gevc}{\ensuremath{{\mathrm{\,Ge\kern -0.1em V\!/}c}}\xspace}
\newcommand{\mevc}{\ensuremath{{\mathrm{\,Me\kern -0.1em V\!/}c}}\xspace}
\newcommand{\gevcc}{\ensuremath{{\mathrm{\,Ge\kern -0.1em V\!/}c^2}}\xspace}
\newcommand{\mevcc}{\ensuremath{{\mathrm{\,Me\kern -0.1em V\!/}c^2}}\xspace}

\def\mus  {\ensuremath{\rm \,\mus}\xspace}

\def\mus        {\ensuremath{\,\mu{\rm s}}\xspace}    

\def\to                 {\ensuremath{\rightarrow}\xspace}

\def\pep2{PEP-II}

\def\gsim{{~\raise.15em\hbox{$>$}\kern-.85em
          \lower.35em\hbox{$\sim$}~}\xspace}
\def\lsim{{~\raise.15em\hbox{$<$}\kern-.85em
          \lower.35em\hbox{$\sim$}~}\xspace}

\def\Vtd  {\ensuremath{|V_{td}|}\xspace}

\def\Vub  {\ensuremath{|V_{ub}|}\xspace}
\def\Vcb  {\ensuremath{|V_{cb}|}\xspace}

\xspace

\def\jetset74   {\mbox{\tt Jetset \hspace{-0.5em}7.\hspace{-0.2em}4}\xspace}

\newcommand{\slideframe}[1]{{}}

\newcommand{\bei}{\begin{itemize}}
\newcommand{\eei}{\end{itemize}}

\def\OMIT#1{{}}

\newcommand{\LQCD}{{\Lambda_{\rm QCD}}}
\newcommand{\texthalf}{{\textstyle{\frac12}}} 
\newcommand\spur{\raise.15ex\hbox{/}\kern-.57em }

\newcommand{\CO}{\mathcal{O}}
\newcommand{\CA}{\mathcal{A}}

\newcommand{\ecut}{{E_{\text{cut}}}}
\def\d{{\rm d}}
\def\FD{{\cal F}}
\def\FDs{\FD_*}

\begin{document}

\Title{Conference Summary}
\bigskip

%
\label{GrinsteinSummaryStart}

%
\author{Benjam\'\i{}n Grinstein}

%
\address{Department of Physics\\
University of California, San Diego\\
La Jolla, CA 92093-0319, USA\\
}

\makeauthor

\section{Flavor Physics Deep Questions}
Having heard of the tremendous progress seen in the last year in the
field of flavor physics and CP violation, I would like to invite the
audience to contemplate some of the questions we are ultimately trying
to answer, and to ponder how much progress has been made towards
answering them. Some of these deep questions are
\begin{itemize}
\item Why are there several generations of quarks and leptons? Are
  there precisely three? Is there some underlying structure that explains
  the presence of generations? Are the generations really identical except for the mass, or
  is there some basic distinction between them?
\item Is there a fundamental explanation to the hierarchy of masses?
  Is there are a relation between quark and lepton masses, or between
  quarks/leptons  of different generations?
\item What gives rise to the texture of mixings between generations?
  Is this related to the masses of quarks and leptons?
\item Is flavor physics tied to Electro-Weak symmetry breaking?
\item What is the origin of the observed CP violation? Is the CP
  violation in the mixings (of quarks and leptons) enough to
  accommodate  baryogenesis or leptogenesis? Else, what is the nature
  of additional CP violation?
\item Is there a relation between quarks and leptons? Are the patterns
  of masses, mixings and CP violation related?
\end{itemize}
The lack of evidence for quark and lepton substructure makes difficult
addressing these questions. It is worth revisiting our current
approach. Seemingly the community of flavor/CP
physicists has concentrated their efforts on measuring with ever
increasing precision the parameters of the standard model that pertain
to flavor physics and CP violation, {\it e.g.}, masses and  
mixing parameters. 

Why is this the right strategy? In the absence of indications of
underlying structure, we are left with the task of looking for ever
subtler deviations from the predictions of the standard model. We
ought to be testing specific theories of flavor, like supersymmetric
extensions of the standard model, extended technicolor, leptoquarks,
overlapping branes, CKM textures, etc. But detailed quantitative tests of these mathematical constructs have  never resulted in confirmation of a theory. What is worse,
rarely is a theory ruled out (a famous counterexample is the
super-weak theory of CP violation). Some of these theoretical
constructs are not very specific in their predictions and are
rightfully largely ignored.  And most models are only marginally
predictive: they share the disturbing feature that they can be made as
unobservable as needed by assuming that their effects become
significant only at ultra-high energies. 

Since theories of flavor fail to indicate useful directions in which we
should test them, we simply press for higher precision and for tests
of consistency of the standard model of flavor: the dominant task in
the field is to both predict and measure at the highest precision
possible. Ever more precise results are parametrized in the CKM
standard model, a theory which does not address flavor, but rather
accommodates it.

It comes as little surprise that most of the talks at this conference
focus on precise determination of parameters and tests of consistency.
While answering the ``deep questions'' remains the primary goal, the
daunting task of determining the (possibly changing) standard is
interesting enough in its own right.  We should succeed at describing
the universe in detail, if not in explaining why the universe is that
way!

Little thought has been given to the scenario in which the standard
CKM model successfully accommodates every result regardless of
precision attained. It may then be that there is no underlying
quantum field theory model of flavor. Instead, the answer to some of
the ``deep questions'' may require thinking outside the box, and may
require invoking some often maligned ideas such as the anthropic
principle and quantum cosmology.

This write-up of my summary talk, as the talk itself,  leaves out many
interesting subjects presented at the conference. I apologize to
those participants whose hard work I could not include here. This is
entirely due to space and time limitations. I have freely quoted from
the talks, and have therefore limited my citations to other entries in
these proceedings, with a few exceptions that correspond to instances
in which I had to consult or quote additional external sources. 

\section{The flavor of leptons: Neutrino masses and mixing}
\begin{figure}
\centering
\includegraphics[height=8.4cm]{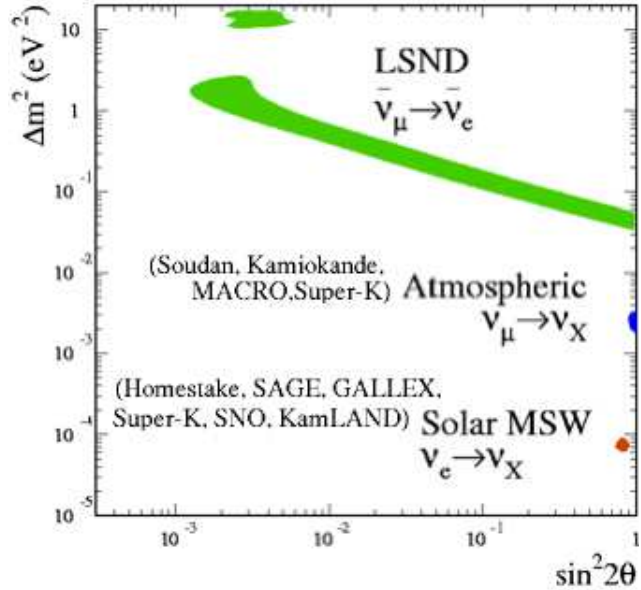}
 \caption{Summary of neutrino mass and mixing evidence\protect\cite{Prebys}.}
\label{fig:neut-mass-mix}
\end{figure}

It is now firmly established that neutrinos mix and therefore $m_\nu
\neq0$. A summary of evidence for neutrino masses and mixings is shown
in Fig.~\ref{fig:neut-mass-mix}. The oscillation probability,
$P(\nu_1\to \nu_2) = \sin^22\theta \sin^2(1.27\Delta m^2L/E)$, depends
both on the mixing angle $\theta$ and the mass difference $\Delta
m$. The atmospheric neutrino data ($\nu_\mu \to \nu_X$) is confirmed
and gives $\Delta m^2\sim 2\times10^{-3}\text{eV}^2$ with near-maximal
mixing. Solar neutrino oscillations ($\nu_e \to \nu_X$) are also
confirmed, with non-maximal mixing and $\Delta m^2\sim
8\times10^{-5}\text{eV}^2$. The LSND reports small angle oscillations
in $\bar \nu_\mu\to \bar \nu_e$ with small mixing angle and $\Delta
m^2\sim0.1$--$10\text{~eV}^2$. The MiniBoonNE experiment will be in a
position to confirm or refute this observation.  Since neutrinos are
neutral their mass may be either Dirac or Majorana.
 
A Dirac mass would make the lepton sector similar to the quark
sector. It requires new fields beyond those in the standard model,
namely, right handed  neutrinos. This raises some interesting
questions.  How many right handed neutrinos are there? How many of them
are active and how many inactive? In this scenario a lepton sector
version of the CKM matrix arises naturally, the
Pontecorvo-Maki-Nakagawa-Sakata matrix (PMNS), and one can ask if this
is in any way related to the CKM matrix and whether the masses are
related to those of quarks. Total lepton number is conserved, just
like baryon number is conserved in the quark sector. And just like for
quarks, individual flavor number is broken. The PMNS automatically
accommodates CP violating phases which give rise to the possibility of 
CP violation for leptogenesis and raises the question of measuring
lepto-CP violation in the lab.

%

A Majorana mass would be even more exciting. It is a new phenomenon
and raises new questions. For example, are there new scalar fields
with vacuum expectation values responsible for these masses? Lepton
number is violated (the mass term has $|\Delta L|=2$). The Majorana
neutrino is by necessity active ($\Delta I_w=1$). The mass could arise from
a triplet Higgs or from the see-saw mechanism through a dimension five
term in the Lagrangian. The latter, being non-renormalizable, could be
interpreted as the result of new short distance interactions. In
addition there could be sterile neutrinos with a bare mass (or with a
singlet Higgs giving mass) which could mix with the active ones. 

The observations indicate an interesting pattern of masses. Ignoring
the LSND observations one has two disparate mass differences
indicating that two neutrinos are almost degenerate, on the scale of
the mass difference with the third. However, since there is no
measurement of the individual mass of any neutrino it is not known
whether the almost degenerate pair is lighter (``normal hierarchy'') or
heavier (``inverted hierarchy'') than the third neutrino. The normal
hierarchy is akin that observed in the quark sector, and it will be
interesting to see if there is a connection between sectors. For a
Majorana mass the inverted hierarchy may lead to observable
neutrino-less double beta decay, but for the normal hierarchy the rate
is unobservably much smaller.
 
The LSND observation, if confirmed, requires the addition of a new
neutrino, since three disparate mass differences require four distinct mass
eigenstates. Measurements of the width of the $Z$-vector boson set to
three the number of active neutrinos with mass less than
$m_Z/2$. Hence any additional light neutrinos must be sterile. Some
constraints on these already exist from present experiments. Pure $\nu_\mu
- \nu_s$ mixing is excluded for atmospheric neutrinos (SK, MACRO), while pure $\nu_e
- \nu_s$ mixing is excluded for solar neutrinos (SK, SNO). The mass
patterns can be of two types, either ``2+2'', in which two pairs of
neutrinos are degenerate on the scale of the large (LSND) mass
difference, or ``3+1'', in which a much heavier neutrino is added to
either of the hierarchies of the previous paragraph.

The future bodes well and busy for neutrino physics. Many experiments
are continuing and many more proposed. MiniBoonNE will either confirm
or refute the LSND findings. Long baseline experiments will narrow the
$\nu_\mu-\nu_\tau $ parameter space, search for $\nu_\mu\to\nu_e$ and improve
significantly the precision of oscillation parameters (K2K-II, MINOS,
CNGS, NOvA, T2K). Proposed dedicated reactor experiments (Angra dos
Reis, Braidwood, Chooz-II, Diablo Canyon, Daya Bay, Kashiwazaki,
Krasnoyarsk) will measure $\theta_{13}$ with higher sensitivity and may
establish it does not vanish\cite{Kobayashi}. Proposed double beta decay
experiments may demonstrate the Majorana nature of neutrino mass
(CUORE, GENIUS, Majorana, super-NEMO, MOON,EXO). Searches for CP
violation in the lepton sector could culminate in a complete theory of
leptogenesis\cite{Kang}.  There is reason to be optimistic!

\section{The CKM Matrix}
\subsection{Status}
Elements of the CKM  matrix, 
\beq 
\vckm=\begin{pmatrix}
V_{ud}&V_{us}&V_{ub}\\
V_{cd}&V_{cs}&V_{cb}\\
V_{td}&V_{ts}&V_{tb}
\end{pmatrix}
\eeqn
are determined with widely varying precision. While unitarity can be
used to significantly constrain individual elements of the matrix,
a determination that does not assume  unitarity tests this
feature. Let's briefly review our knowledge of individual elements,
without assuming unitarity.

%
\begin{figure}
\centering
\includegraphics[height=8.4cm]{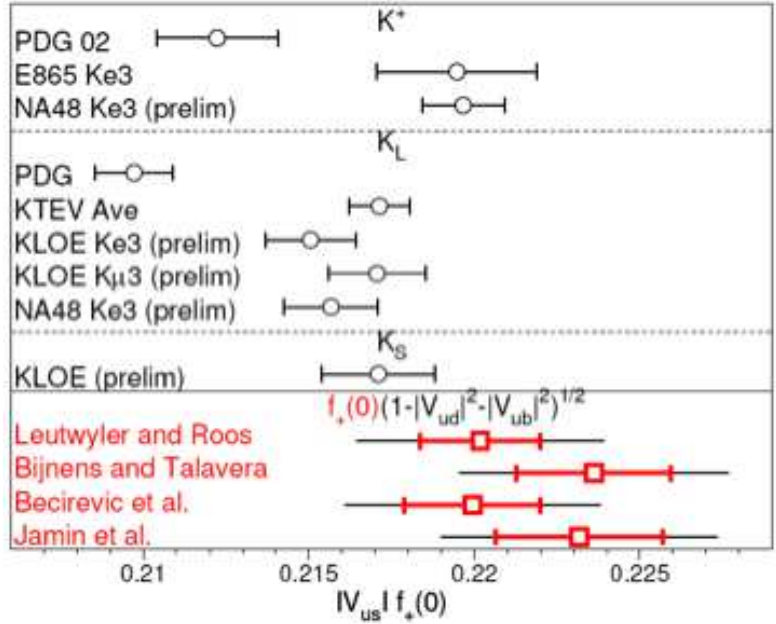}
 \caption{Current status of experiments in the $V_{us}$ determination\protect\cite{Glazov}.}
\label{fig:glazov}
\end{figure}

The best determined element is $V_{ub}$. It is obtained from nuclear
beta decay and the error is only 0.2\%. 

Next best known is the Cabibbo angle, or $V_{us}$, which the PDG
claims to be known at the 1\% level.  It is determined from $K\ell3$
decays. However, there are inconsistencies between the most recent
(E685, NA48, KTeV, KLOE) measurements of decay rates and the summary
of older experiments in the PDG. This is just as well because there
are also inconsistencies between the quoted values and unitarity of
the CKM. A tremendous effort was undertaken to resolve these
inconsistencies\cite{Glazov}. The current experimental situation is
summarized in Fig.~\ref{fig:glazov}, which shows that the new rates
are all consistently higher than the PDG values. The decay rate is
given in terms of a form factor $f_+$ (parametrizing the matrix
element of the hadronic current) and the CKM element, so to extract
$V_{us}$ and test unitarity of the CKM matrix one needs theory to
predict the form factor. The last row in Fig.~\ref{fig:glazov}\ shows
the rate as predicted by CKM-unitarity using different theory
calculations of the form factor. The degree of consistency depends on
which calculation is adopted. Just as with experiment, there is a
renewed effort to re-analyze the theory of these decays. Two new
calculations of form factors have appeared, one using chiral
perturbation theory extending the classic
calculation of Leutwyler and Roos, and one using quenched lattice
QCD. While the calculations differ only by
about 2\%, the deviation is of practical importance, particularly when
the PDG claims a precision of 1\% in $V_{us}$. It is tempting to favor
the lattice calculation over the chiral perturbation theory one,
partly because it agrees with the older Leutwyler-Roos and partly
because the lattice is supposed to be unadulterated QCD. But, reader
beware, quenched lattice calculations are non-systematic
approximations. Moreover, the deviation between theoretical
calculations is as large as one could expect it to be! Recall that
the form factor is protected by the Ademollo-Gatto theorem: \beq
f_+(0)=1+0\cdot\epsilon+\CO(\epsilon^2) \eeqn where $\epsilon\sim m_s/
\LQCD\sim1/3$ is the dimensionless symmetry breaking parameter
($\epsilon\sim m_K^2/ \Lambda_\chi^2\sim1/4$ in chiral perturbation
theory). If the quenching error is of typical magnitude, $\sim20$\%,
then the error in $f_+(0)$ ought to be $20\text{\%}\times
\epsilon^2\sim2\text{\%}$.

It is rather surprising that $V_{cd}$ and $V_{cs}$ are poorly
known. Charm production in neutrino scattering ($\nu+ d\to c+X$) gives a
5\% determination of $V_{cd}$, while $W^+ $ decays to charm give a 10\%
determination of $V_{cs}$. These are two elements where progress would
be welcome. It is hard to believe that these could not benefit from
the advances in understanding of inclusive semileptonic decays of
heavy mesons, which have yielded a $1-2\text{\%}$ measurement of
$V_{cb}$. 

Currently, the errors in the determination of $V_{cb}$ and $V_{ub}$
are about 1\%\ and 10\%, respectively. These will be discussed in some
detail below, since they are the focus of  a large effort in  the
flavor community and there has been much progress recently. 

The last row of the CKM is known the least. $V_{tb}$ is determined at
the 20\%\ level from top quark decay, but $V_{td}$  and $V_{ts}$ can
only be accessed indirectly, through electroweak loops as in, for
example, $B^0-\bar B{}^0$ mixing and radiative $B$ decays.

\section{Over-constraining the unitarity triangle}
Unitarity of the CKM matrix and the freedom to redefine fields result
in only four independent parameters needed to parametrize the
matrix. The Wolfenstein parametrization, 
\beq 
\vckm=\begin{pmatrix}
V_{ud}&V_{us}&V_{ub}\\
V_{cd}&V_{cs}&V_{cb}\\
V_{td}&V_{ts}&V_{tb}
\end{pmatrix}
\approx \begin{pmatrix}
1-\texthalf\lambda^2&\lambda&A\lambda^3(\bar\rho-i\bar\eta)\\
-\lambda(1+iA^2\lambda^4\bar\eta & 1-\texthalf\lambda^2& A\lambda^2\\
 A\lambda^3(1-\bar\rho-i\bar\eta)& -A\lambda^2(1+i\lambda^2\bar\eta)& 1
\end{pmatrix} + \mathcal{O}(\lambda^6)
\eeqn
\begin{minipage}[b]{0.6\linewidth}
shows the four parameters explicitly and is  most frequently
used.
Joining the points 0, 1 and $\bar\rho+i\bar\eta$ in the complex plane
yields a {\sl unitarity triangle.} The sides have length 1,
$\sqrt{\bar\rho^2+\bar\eta^2}=|V_{ub}|/|V_{us}V_{cb}|$ and
$\sqrt{(1-\bar\rho)^2+\bar\eta^2}=|V_{td}|/|V_{us}V_{cb}|$. The triangle is
determined, up to discreet ambiguities, by these lengths or,
alternatively, by two of the three vertex angles, defined in the
figure to the right.
 If the CKM
picture of flavor and CP physics is correct, the same triangle should
be inferred from decay rates as from CP asymmetries.
\end{minipage}
\begin{minipage}[b]{0.5\linewidth}
\center
\epsfig{file=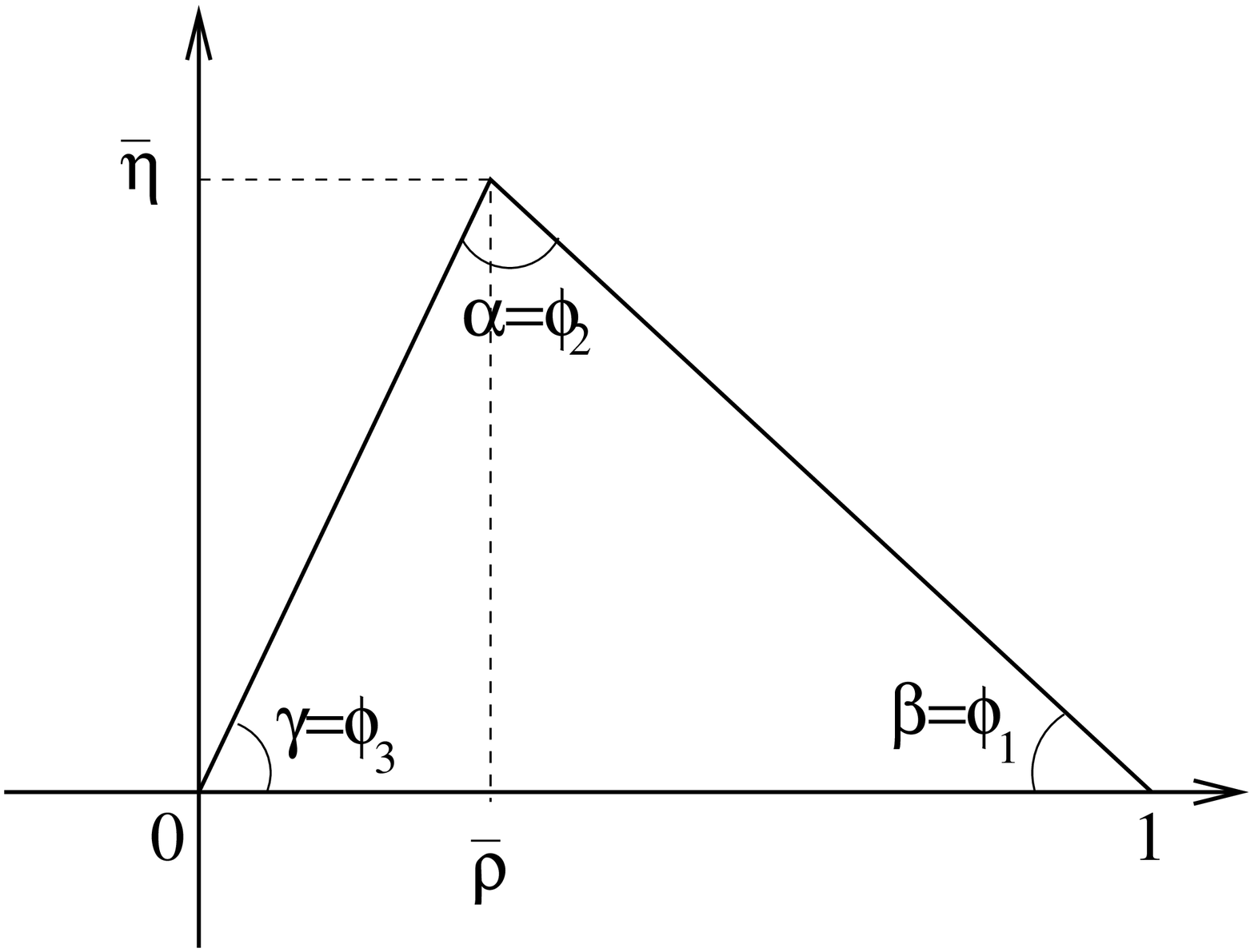,width=5.5cm}
\end{minipage}

%
\begin{figure}
\centering
\includegraphics[height=8.4cm]{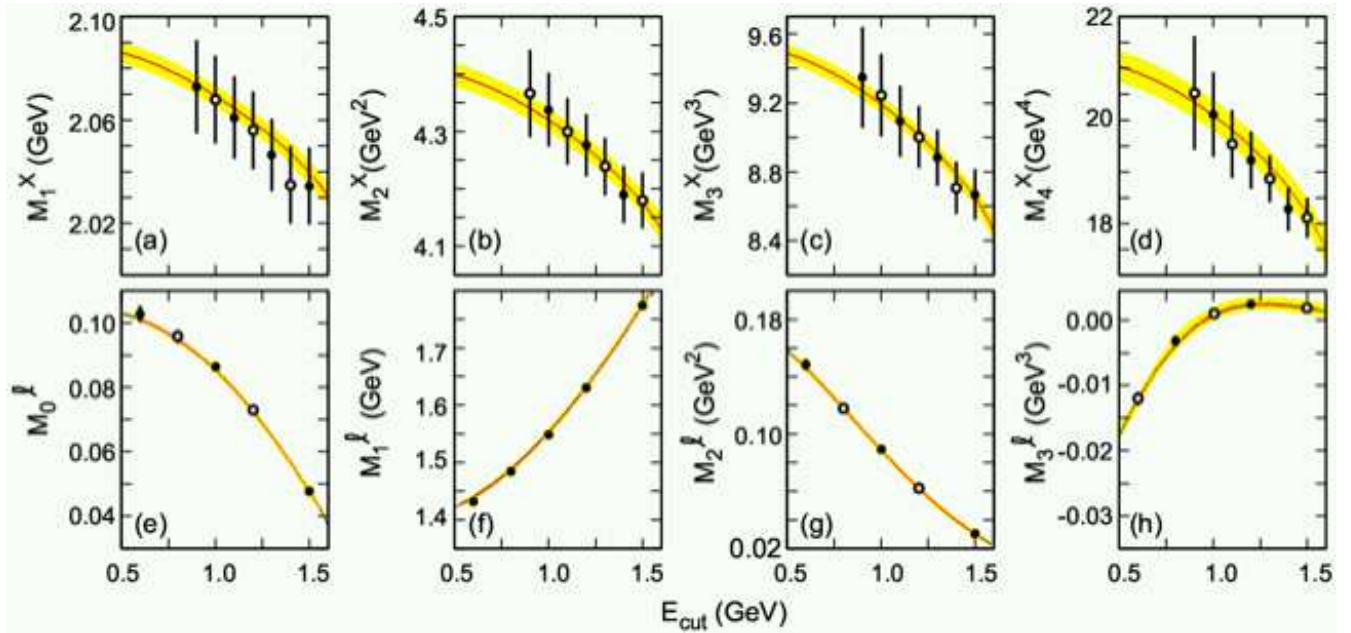}
 \caption{Babar fit to moments of the $B\to X_c\ell\nu_\ell $ spectrum\protect\cite{Barberio}.}
\label{fig:babar-fit}
\end{figure}

\subsection{Sides}
\subsubsection{$|V_{cb}|$ from inclusive semileptonic decays}
In order to calculate the rate $d\Gamma(B\to X_c\ell\nu_\ell )$ theorists use an
expansion in powers of $\LQCD/m_b$. The latest analyses retain terms
up to and including order $(\LQCD/m_b)^3$, incurring in an error
$\sim(\LQCD/m_b)^4\sim(0.1)^4$. The expansion introduces unknown
non-perturbative parameters. These can be determined by measuring 
the semileptonic spectrum precisely. 

To this effect experiment measures, and theory predicts, moments of the
the decay spectrum. Commonly used are lepton energy moments,
\beq
\vev{E^n_\ell}_\ecut\equiv
\frac{R_n(\ecut,0)}{R_0(\ecut,0)}\quad\text{where}\quad
R_n(\ecut,M)\equiv \int_\ecut\!\!\!(E_\ell-M)^n
\frac{\d\Gamma}{\d E_\ell}\,\d E_\ell\,,
\eeqn
and hadronic mass moments
\beq
\vev{m_X^{2n}}_\ecut\equiv
\frac{\int_\ecut\!(m_X^2)^n\frac{\d\Gamma}{\d m^2_X}\,\d m^2_X}%
{\int_\ecut\!\frac{\d\Gamma}{\d m^2_X}\,\d m^2_X}\,.
\eeqn
where $m_X^2\equiv (p_B-p_\ell-p_\nu)^2=(p_B-q)^2$. Note that the moments are
defined with a lepton energy cut. While this cut is an experimental
necessity, measuring moments as a function of the cut allows for a
stringent test of the theory and a better determination of
parameters. Also included in the
analysis are photon energy moments in $d\Gamma(B\to X_s\gamma)$, which depend
on the same unknown hadronic parameters as their semileptonic
counterparts.

BaBar has performed a fit\cite{Barberio} using the theoretical analysis of
Gambino and collaborators and of Bataglia and Uraltsev. The results, in
Fig.~\ref{fig:babar-fit}, show very nice agreement between theory and
experiment, and the resulting value for the CKM angle is
$|V_{cb}| = (41.4\pm0.4_{\text{exp}}\pm0.4_{\text{HQE}}\pm0.6_{\text{theory}})\times10^{-3}$.
In addition, Bauer and collaborators have performed a global fit to data from
BaBar, Belle, CDF, CLEO and DELPHI, with comparable results, as
explained elsewhere in these proceedings\cite{bg}.

%
\begin{figure}
\centering
\includegraphics[height=5.4cm]{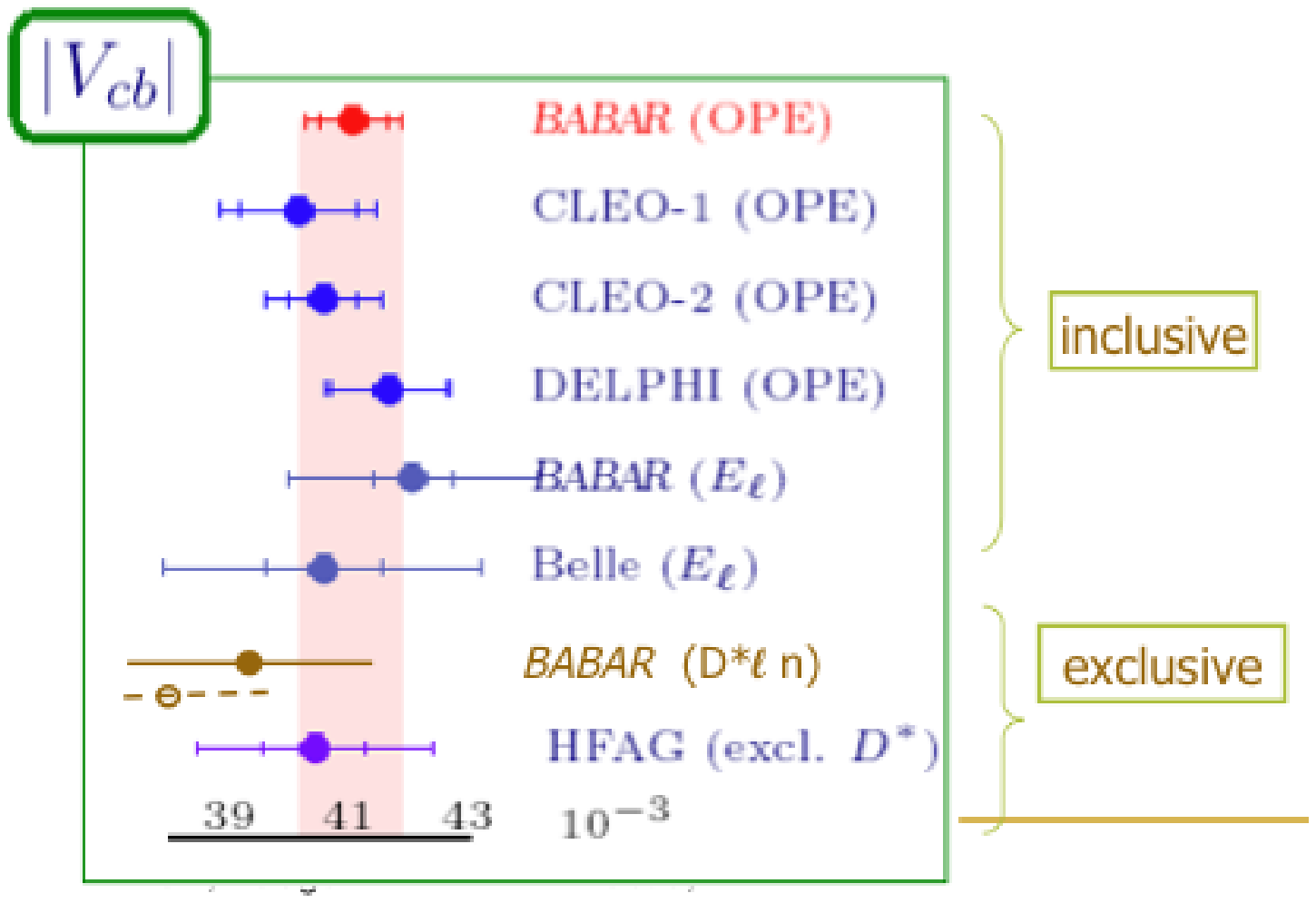}
\hspace{1cm}\includegraphics[height=6.0cm]{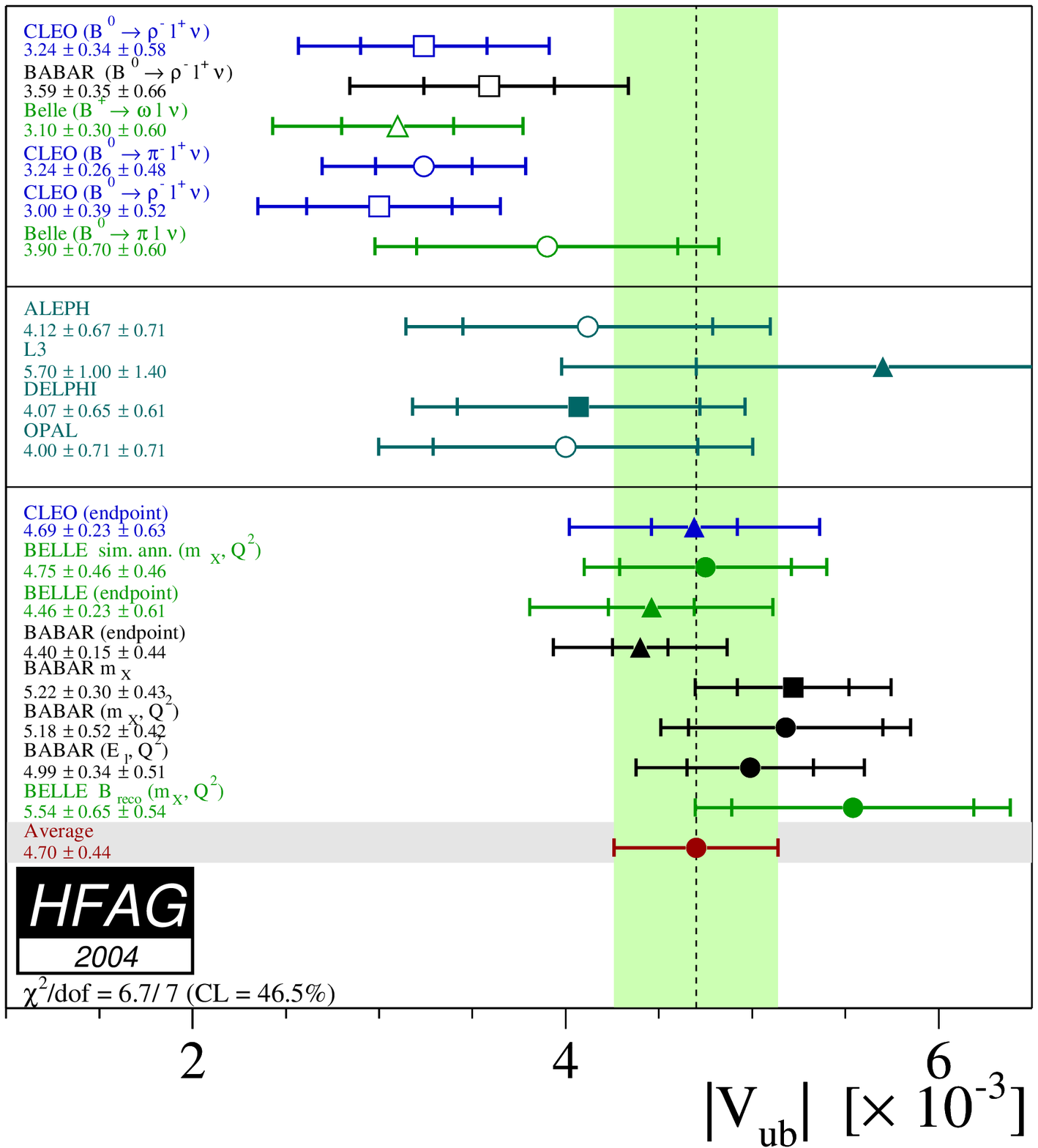}
\caption{Summary of $\Vcb$ and $\Vub$ determinations.}
\label{fig:vcb-vub}
\end{figure}

\subsubsection{$|V_{cb}|$ from exclusive semileptonic decays}
The differential decay rate
\begin{equation*}
  \frac{\d\Gamma(B\to D^*\ell\bar\nu)}{ \d w} = \frac{G_F^2 m_B^5}{ 48\pi^3}\,
  r_*^3\, (1-r_*)^2\, \sqrt{w^2-1}\, (w+1)^2 \left[ 1 + \frac{4w}{
  1+w} \frac{1-2wr_*+r_*^2}{ (1-r_*)^2} \right] |V_{cb}|^2\,{
  \FDs{}^2(w)}
\end{equation*}
is parametrized in terms of $\FDs$, a combination of form factors of
the $V-A$ charged current. Here $w\equiv p_B\cdot p_D^{*}/M_BM_D^{*}$, and
$r^{*}=M_D^{*}/M_B$. At lowest order in HQET $ \FDs(1)=1$, and Luke's
Theorem insures the corrections are small,
$\FDs(1)-\eta_{\text{QCD}}=\mathcal{O}(\LQCD/m_c)^2$, where
$\eta_{\text{QCD}}$ is a known short distance QCD correction. Since the
rate vanishes at $w=1$, to determine $|V_{cb}|$ an extrapolation of
the data to $w=1$ must be made. This is aided by theory, since
analyticity and unitarity tightly constrain the functional form of ${
\FDs(w)}$.

The precision in the determination of $|V_{cb}|$ from exclusive decays
is limited by a 4\%\ uncertainty in $\FDs(1)$. While
only marginally  competitive with the inclusive determination, it
provides an independent test of theory.

\subsubsection{$|V_{ub}|$ from inclusive semileptonic decays}
The rate for charm-less semileptonic $B$ decays is much smaller than
for charm-full decays. Hence experimental kinematic cuts that exclude
the charm-full decays are needed. One may cut in the hadronic mass,
$m_X^2=(P_B-p_\ell-p_\nu)^2$, the lepton mass $q^2=(p_\ell+p_\nu)^2$
or the charged lepton energy $E_\ell$. In all three cases the rate is
significantly limited, with the cut on $q^2$ being the most
limiting. The theory of inclusive charm-less $B$ decays is, in
principle, the same as for charm-full decays. But in practice the
necessary cuts complicate matters. If the cut is not too stringent so the
decay is not dominated by a few resonances but rather by low invariant
mass jets then the theory can still be organized through an OPE but
now an infinite number of ``leading twist'' terms contribute
equally. This infinite sum can be parametrized by a shape
function which encodes our ignorance of strong interactions. The
challenge is to make model independent predictions. 

At leading order in a $1/m_b$ expansion the shape function in $B\to
X_u\ell\nu$ is the same that appears in the rate for $B\to
X_s\gamma$. In principle then one can determine the shape function
from $B\to X_s\gamma$ and use it in the determination of $|V_{ub}|$ from $B\to
X_u\ell\nu$. In practice, however, the $1/m_b$ corrections, which
spoil the equality of the shape function in the two processes, can be
large. The naive guess $\LQCD/m_b\sim10\text{\%}$ is not far off
from detailed estimates. 

Both Belle and BaBar have determined $|V_{ub}|$ by this method. I quote
here the result but warn the audience that the theoretical
uncertainties from $1/m_b$ corrections have not been properly
accounted for resulting in a (probably large) underestimate of
theoretical error:
\begin{align*}
\text{Belle:}\quad\Vub &=  (5.54 \pm 0.42 \pm 0.55 \pm 0.42 \pm 0.27) \times 10^{-3}\\
\text{BaBar:}\quad\Vub &= (5.18 \pm 0.41 \pm 0.40 \pm 0.23 \pm 0.27) \times 10^{-3}
\end{align*}
Here the errors are statistical, experimental, fit and OPE (theory),
respectively. Fig.~\ref{fig:vcb-vub} shows the HFAG compilation of results
for the determination of $|V_{ub}|$\cite{hfag}.

\begin{figure}
\centering
  \includegraphics[height=6.0cm]{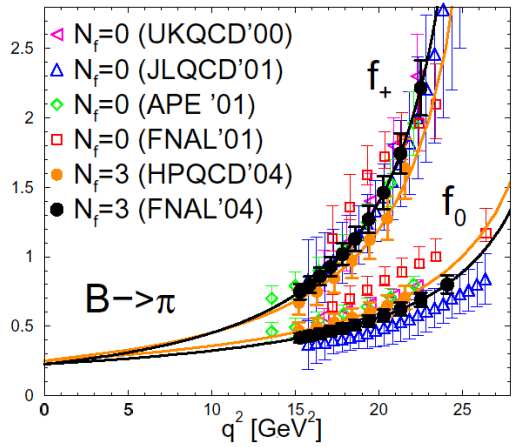}
\caption{Form factors for $B\to \pi\ell\nu$ computed by simulations of
  lattice QCD\cite{okamoto}. }
\label{fig:B->pi}
\end{figure}

\subsubsection{$|V_{ub}|$ from exclusive semileptonic decays}
As in the case of inclusives the situation for $|V_{ub}|$ is worse than
for $|V_{cb}|$. The problem is that HQET does {\it not} fix
normalization at zero recoil as it does in $B\to D^{(*)}\ell\nu$. For $B\to\pi\ell\nu $
two form factors determine the decay amplitude: The determination of
$\Vub$ from exclusive decays requires {\it a priori} knowledge of the
form factor $f_+$ for $B\to\pi$, defined through
\beq
\vev{\pi(p_\pi)|\bar b\gamma_\mu q| B(p_B)}
= \left((p_B+p_\pi)_\mu-\frac{m_B^2-m_\pi^2}{q^2}q_\mu\right){ F_+(q^2)}
+\frac{m_B^2-m_\pi^2}{q^2}q_\mu { F_0(q^2)}\,,
\eeqn
where $q\equiv p_B-p_\pi $. Analyticity and unitarity do constrain the
functional form, which helps, {\it e.g.,}  to interpolate lattice results.
A compilation of lattice results is shown in
Fig.~\ref{fig:B->pi}. One can use this to determine $\Vub$, but the
errors are large. Belle finds, from $B\to\pi\ell\nu$ restricted to
$q^2>16\text{GeV}^2$, $\Vub=(3.87\pm0.70\pm0.22 ^{+0.85}_{-0.51}) \times
10^{-3}$ and $\Vub= (4.73 \pm 0.85 \pm 0.27 ^{+0.74}_{-0.50}) \times 10^{-3}$
using lattice results from FNAL'04 and HPQCD,
respectively.
The theory errors,  
$\sim \pm20$\%, are not expected to be significantly reduced soon. There has
also been some effort to understand the precision with which $\Vtd$
can be determined from $B\to\rho\gamma$ decays\cite{bg}.

New ideas are needed to reduce the error on $\Vub$ from exclusive
decays to the sub-10\% level, hopefully to a few percent. One recent
proposal is to use double ratios
to eliminate hadronic uncertainties. While the method is promising a
detail theoretical error analysis is missing, as are experimental
feasibility studies\cite{bg}.

\subsection{Angles}
The direct route to measuring angles is through CP violation in
interference between decay and mixing in the decay of a neutral $B$
meson to a CP eigenstate $f_{\text{CP}}$. The asymmetry
\beq
a_{f}=\frac{\Gamma[\overline B{}^0(t)\to f]-\Gamma[B^0(t)\to
    f]}{\Gamma[\overline B{}^0(t)\to f]+\Gamma[B^0(t)\to f]} =S_f\sin(\Delta m t)-C_f\cos(\Delta m t)
\eeqn
is theoretically clean if the final state is a CP eigenstate and only
one weak phase contributes (or is dominant), in which case
\beq
a_{f_{\text{CP}}}=\text{Im}\;\lambda_{f_{\text{CP}}} \sin(\Delta m t).
\eeqn
Here $\lambda_{f_{\text{CP}}}=(q/p)\bar \CA_{f_{\text{CP}}}/
\CA_{f_{\text{CP}}}$ depends on the weak mixing angles through the
$B^0-\overline B{}^0$ mixing parameters $p$ and $q$, and through the
ratio of amplitudes for $\overline B{}^0$ and $B^0$ decays,
$\overline\CA_{f_{\text{CP}}}$ and
$\CA_{f_{\text{CP}}}$, to the common final state $f_{\text{CP}}$. The
coefficient of $\cos(\Delta mt)$ is also often denoted by  $A_f=-C_f$, but
this notation may be confusing given the proliferation of A's (in Wolfenstein's
parametrization, to denote asymmetries and to denote amplitudes).

\begin{figure}
\centering
  \includegraphics[height=6.0cm]{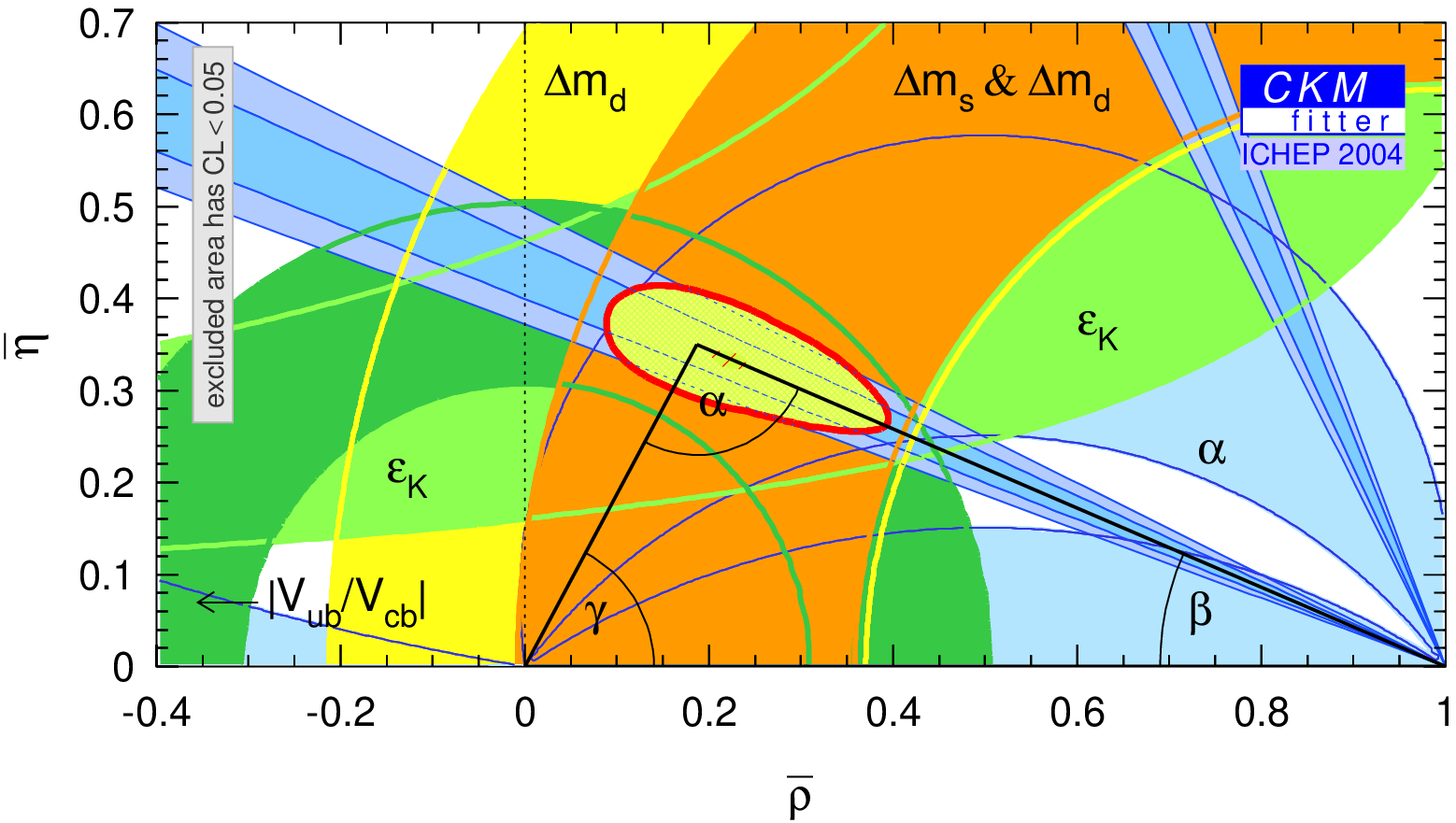}
  \includegraphics[height=6.0cm]{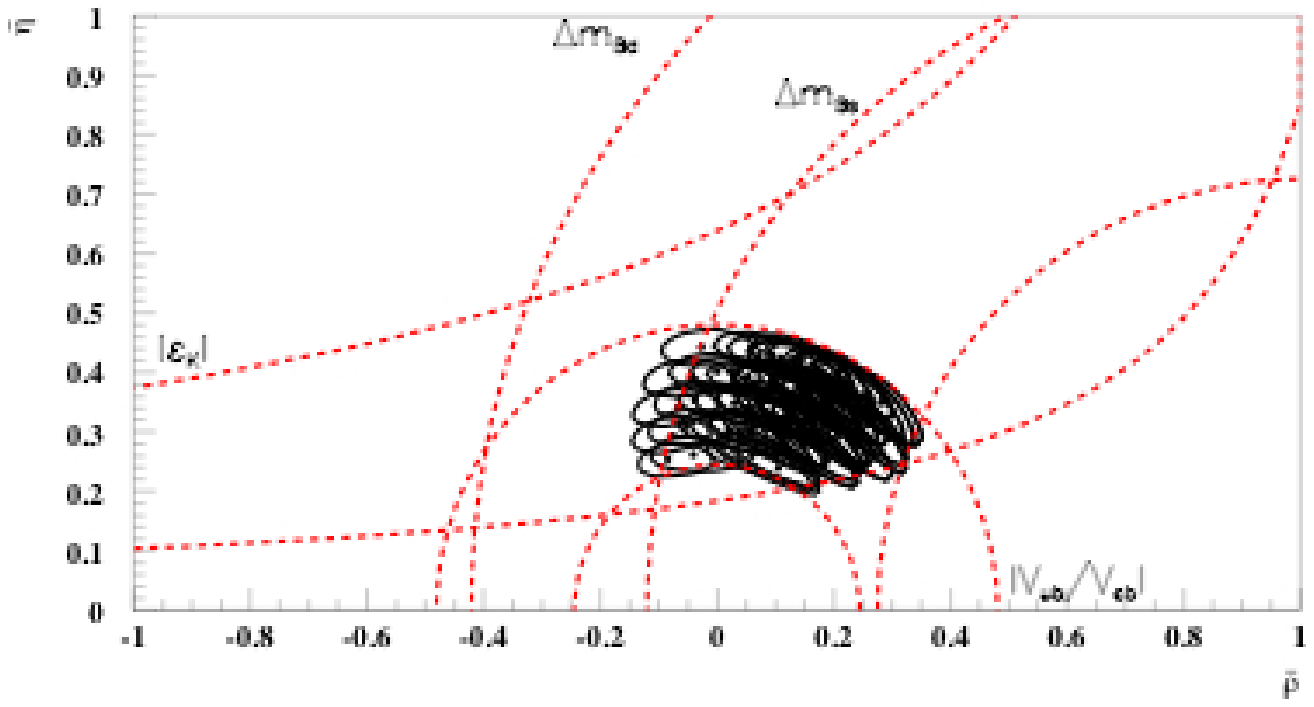}
\caption{Standard model fit as reported by
  CKMfitter\protect\cite{ckmfitter} (above) and in   the BaBar book,
  circa 1998 (below). The figures are drawn to different scale.
  }
\label{fig:ckmfitter}
\end{figure}

\subsubsection{$\sin(2\phi_1)$ from $b\to c\bar cs$}
Great precision has been achieved by both Belle and BaBar in the
determination of $\sin(2\phi_1)$ from $B$ decays to charmonium plus
$K^0$\cite{Trabelsi}:
\begin{align*}
\text{BaBar:}\quad\sin(2\phi_1) &=+0.722 \pm 040\pm 023\\
\text{Belle:}\quad\sin(2\phi_1) &=+0.728 \pm 056\pm 033
\end{align*}
We can finally over-constrain the unitarity
triangle. Figure~\ref{fig:ckmfitter} shows CKMfitter standard model fit
including this $S_{\psi  K}$ data.  The obvious first remark is that
the standard model works remarkably well. Clearly we want to keep piling on
observables to this global fit. It should be noted that the
determination of $\cos(2\phi_1)$, while interesting, adds little to this
program.  It is also quite clear, from the precision achieved, that
further progress requires we concentrate on clean observables, for
which model dependence does not cloud the issue of interpretation of
experimental results.

It is interesting to ponder on what the best next direction may
be. From Fig.~\ref{fig:ckmfitter}\ it is clear that pinning down the
apex of the unitarity triangle would be best achieved by measuring the
length of the side opposite the origin, that is, $|1-(\bar \rho +i\bar
\eta)|$, since this is orthogonal to the direction fixed by $\phi_1$. This
is tested by $B^0-\bar B{}^0$ oscillations and by $b\to d$ decays, like
$B\to \rho\gamma$ (for which a branching fraction upper limit of $1.2\times10^{-6}$
has been established). Alternatively, one may measure 
$\sin(2\phi_3)$ with precision.

But if testing consistency of the picture is what we are after,
measuring the length $|\bar \rho +i\bar \eta|$ is best, since this is
largely parallel to the direction fixed by $\phi_1$. Hence advances in
measuring $|V_{ub}|$ are of paramount importance.

It is nice to see how much progress has been made. Figure~\ref{fig:ckmfitter},
reproduced from the BaBar book (circa 1998), shows the status in the
determination of the unitarity triangle six years ago, allowing  a much
wider region  for the apex of the triangle.

\subsubsection{$\sin(2\phi_1)$ from $b\to c\bar cd$, $b\to q\bar qs$}
$\sin(2\phi_1)$ can also be determined, albeit less cleanly, from
$b\to c\bar cd$, $b\to q\bar qs$ decays. Some extensions of the
standard model allow for significant deviations in $-\eta_f S_f$ from
$\sin(2\phi_1)$ in individual processes (here $\eta_f$ is the CP
parity of the final state $f$). Unfortunately,  theoretical
predictions for these decays are less clean, with an uncertainty in
$-\eta_f S_f-\sin(2\phi_1)$ guessed at~0.1 -- 0.2. 

\begin{figure}
\centering
  \includegraphics[height=6.0cm]{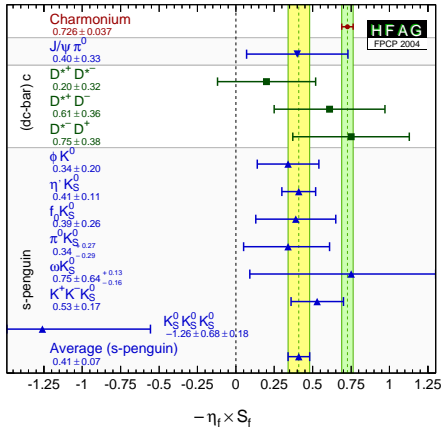}
\caption{Compilation of results for $-\eta_fS_f$ from different
  modes\protect\cite{hfag}. The $K^0_sK^0_sK^0_s$ was new at this conference. }
\label{fig:sin2b_modes}
\end{figure}

Figure~\ref{fig:sin2b_modes} shows a compilation of results for
several processes. The result for $f=K^0_sK^0_sK^0_s$ was first
presented at this conference\cite{Gershon,Hazumi} and shows a central value that
dramatically deviates from expectation albeit with large errors. While
no single observation significantly deviates from expectations, there
is a disturbing (welcome, perhaps?)  trend in the $b\to q\bar q s$
decays. Averaging results,
\begin{align*}
S_{\psi K}-\vev{-\eta_f S_{f(b\to s)}}&=0.30\pm0.08\\
S_{\psi K}-\vev{-\eta_f S_{\eta'K_s,\phi K_s}}&=0.33\pm0.11
\end{align*}
These are $3.5\sigma$ and $3.1\sigma$ effects respectively. It will
be interesting to see how this measurements evolve in the future and
whether theory provides a more certain prediction of the expected
deviation.

These deviations, if they persist, could be accommodated by
non-extravagant extensions of the standard model. This would hardly be
the case if the deviations were much larger.  Take for example SUSY
models\cite{Yoshikawa,Kwon,Ko,rare2}.  While LR insertions are
severely constrained by $B\to X_s\gamma$, 
LL/RR penguins can give significant contributions. These, however, are
beginning to be constrained by $\text{Br}(B\to X_s\ell\ell)=
(4.5\pm1.0)\times10^{-6}$. This is another story whose development is
worth watching!

\subsubsection{Measurement of $\phi_2$ and $\phi_3$}
Another remarkable recent development is the measurement of the other two
unitarity angles, $\phi_2$ and
$\phi_3$\cite{Gershon,Bevan,Clark,Albert,Hazumi,Zupan}.  

Three methods have been used for the measurement of $\sin(2\phi_2)$:
\begin{itemize}
\item $B\to\pi\pi$. This requires an isospin analysis. The branching
  fraction into $\pi^0\pi^0$ and $C_{\pi^0\pi^0}$ have been measured
  making the analysis viable. The results from Belle and BaBar for both
  $\sin(2\phi_2^{\text{eff}})$ and $C_{\pi^+\pi^-}$ differ significantly but
  the errors are still large (for a discussion see \cite{Sanda}).
\item $B\to\rho\rho$ benefits from the small neutral branching fraction,
  $\text{Br}(\rho^0\rho^0)/
  \text{Br}(\rho^+\rho^-)<0.04$\break $(90\text{\%~CL})$. This implies that
  $\phi_2-\phi_2^{\text{eff}} $ is small.
\item $B\to\rho\pi $ Dalitz plot. A clean determination requires a pentagon
  analysis which needs the branching fractions and CP asymmetries for
  the $\rho^+\pi^- $, $\rho^0\pi^0 $ and $\rho^-\pi^0 $ modes. Alternatively, $SU(3)$
  flavor symmetry and factorization of the weak decay amplitude
  assumptions are used.
\end{itemize}
From the combined $\pi\pi$, $\rho\rho$ and $\rho\pi $ analysis,
$\phi_2=(100\begin{smallmatrix}+12\\-11\end{smallmatrix})^\circ$. This
compares rather well with the CKM indirect constraint fit, $\phi_2=(98\pm16)^\circ$.

$\sin(2\phi_3)$ is determined from the interference between $B^-\to
D^0K^-$ and $B^-\to \overbar D{}^0K^-$, achieved with a final state
common to $D^0$ and 
$\overbar D{}^0$. A difficulty here is that the CP
asymmetry is suppressed when the $D^0$ and $\overbar D{}^0$ decay
amplitudes to a particular final state differ vastly. Both $D^0$ and
$\overbar D{}^0$ have Cabibbo allowed decays to $K^0_s\pi^+\pi^-$, and the
best present determination of $\sin(2\phi_3)$ is from that analysis,
\begin{align*}
\text{Belle:}\quad  \phi_3 &= (68 \begin{smallmatrix}+14\\-15\end{smallmatrix}
  \pm 13\pm 11)^\circ~, \\
\text{BaBar:}\quad  \gamma  &= (88 \pm 41  \pm 19\pm 10)^\circ ~,
\end{align*}
where the errors are statistical, systematic and from model
dependence. The much larger error quoted by BaBar is due to the large
correlation in the error in $\gamma$ and the value of $r_B\equiv A(B^-\to
\overline D{}^0K^-)/A(B^-\to D^0K^-)$, which is measured 
by Belle, $r_B=0.26
\begin{smallmatrix}+0.11\\-0.15\end{smallmatrix}\pm0.03\pm0.04$, while
  BaBar only obtains an upper bound $r_B<0.18~(\text{90\%~CL})$. 

\subsubsection{Direct CP violation: The last nail on the superweak
  coffin!}
The superweak theory has all sources of CP violation in $\Delta B=2$ or $\Delta
S=2$ interactions. Direct CP violation in decays of $B$ mesons proceed
via $\Delta B=1$ interactions only, so a positive signal rules out the
superweak theory. However, the interpretation of a signal does not
immediately translate into a measurement of CKM phases. The CP
asymmetry from two interfering amplitudes,
\beq
A_{CP}=\frac{2\sin(\varphi_1-\varphi_2)\sin(\delta_1-\delta_2)}{|A_1/A_2|+|A_2/A_1|+\cos(\varphi_1-\varphi_2)\cos(\delta_1-\delta_2)}
\eeqn
depends on unknown strong interaction phases $\delta_i$ and amplitudes
$A_i$, in addition to the sought after weak phases $\varphi_i$. The combined
analysis of Belle and BaBar gives $A_{CP}(B^0\to K^+\pi^-)=-0.114\pm0.020$,
which is $5.7\sigma$ away from zero\cite{Graziani}. Interestingly, it is
also found that $A_{CP}(B^+\to K^+\pi^0)=-0.049\pm0.040$, which, although compatible with
zero, is $3.6\sigma$ away from  $A_{CP}(B^0\to K^+\pi^-)$ indicating that
color-allowed tree amplitudes do not dominate color-suppressed trees
(plus electroweak penguins). 

\section{Summary of the Summary}
We have seen an evolution in flavor physics since the discovery of $D$
and $B$ mesons from a science which established new important
qualitative facts, such as the long lifetime of $B$ mesons and the
near diagonal structure of the CKM matrix, to what is now a precision
science. At the same time we are witnessing the start of what
promises to be a similar story in the lepton sector: the PMNS matrix
is imprecisely known, much like CKM was 20 years ago. Both camps are
tooling up for a next round of precision improvement and  the future
seems bright\cite{Iijima, Menzemer, Dornan}. 

Have we seen a break with the standard paradigm? Certainly the
positive result for neutrino masses requires some new physics beyond
the standard model, be it new right handed fields or new
interactions. And perhaps the hints of anomalies in $b\to s$ decays are
an indication of surprises to come. Important  steps toward
answering the deep questions!

\section*{Acknowledgments}
I would like to thank the local organizers of the conference for their hospitality and for giving me the opportunity to give this summary talk.
This work is supported in part by a
grant from the Department of Energy under Grant DE-FG03-97ER40546.

%
\label{GrinsteinSummaryEnd}


\begin{thebibliography}{99}

\bibitem{Prebys}Eric Prebys, {\it Status of MiniBooNE}, talk at ICHEP2004


\bibitem{Kobayashi} Takashi Kobayashi, these proceedings, 
  pp.~\pageref{KobayashiStart}--\pageref{KobayashiEnd}

\bibitem{Kang}Sin Kyu Kang, these proceedings, 
  pp.~\pageref{KangStart}--\pageref{KangEnd}

\bibitem{Glazov} Sasha Glazov, these proceedings,  pp.~\pageref{GlazovStart}--\pageref{GlazovEnd}


\bibitem{Barberio}  Elisabetta Barberio, these proceedings,  pp.~\pageref{BarberioStart}--\pageref{BarberioEnd}


\bibitem{bg} Benjamin Grinstein, these proceedings,  pp.\pageref{GrinsteinStart}--\pageref{GrinsteinEnd}


\bibitem{babarbook}
P.~F.~Harrison and H.~R.~Quinn,
``The BaBar physics book: Physics at an asymmetric B factory,''
SLAC-R-0504

\bibitem{hfag}http://www.slac.stanford.edu/xorg/hfag/semi/summer04/summer04.shtml

\bibitem{okamoto} Masataka Okamoto, these proceedings, pp.~\pageref{OkamotoStart}--\pageref{OkamotoEnd}

\bibitem{Trabelsi} Karim Trabelsi, these proceedings, pp.~\pageref{TrabelsiStart}--\pageref{TrabelsiEnd}

\bibitem{ckmfitter}http://www.slac.stanford.edu/xorg/ckmfitter

\bibitem{Yoshikawa} Tadashi Yoshikawa,  these proceedings,
  pp.\pageref{YoshikawaStart}--\pageref{YoshikawaEnd}

\bibitem{Kwon} Youngjoon Kwon,  these proceedings,
  pp.\pageref{KwonStart}--\pageref{KwonEnd}

\bibitem{Ko} Pyungwon Ko, these proceedings,
pp.\pageref{KoStart}--\pageref{KoEnd}

\bibitem{rare2} Alex Lenz,  these proceedings pp. \pageref{LenzStart}--\pageref{LenzEnd}

\bibitem{Gershon} Timothy Gershon, these proceedings,
pp.\pageref{GershonStart}--\pageref{GershonEnd}


\bibitem{Hazumi} Masashi Hazumi, these proceedings,
pp.\pageref{HazumiStart}--\pageref{HazumiEnd}

\bibitem{Bevan} Adrian Bevan, these proceedings,
pp.\pageref{BevanStart}--\pageref{BevanEnd}

\bibitem{Clark} Phil Clark, these proceedings,
pp.\pageref{ClarkStart}--\pageref{ClarkEnd}

\bibitem{Albert} Justin Albert, these proceedings,
pp.\pageref{AlbertStart}--\pageref{AlbertEnd}

\bibitem{Zupan}Jure Zupan, these proceedings,
pp.\pageref{ZupanStart}--\pageref{ZupanEnd}

\bibitem{Graziani} Giacomo Graziani, these proceedings,
pp.\pageref{GrazianiStart}--\pageref{GrazianiEnd}

\bibitem{Sanda} A.\ Ichiro Sanda, these proceedings,
pp.\pageref{SandaStart}--\pageref{SandaEnd}

\bibitem{Iijima}  Toru Iijima, these proceedings,
pp.\pageref{IjimaStart}--\pageref{IjimaEnd}

\bibitem{Menzemer} Stephanie Menzemer, these proceedings,
pp.\pageref{MenzemerStart}--\pageref{MenzemerEnd}

\bibitem{Dornan} Peter Dornan, these proceedings,
pp.\pageref{DornanStart}--\pageref{DornanEnd}


\end{thebibliography}
\end{document}